\documentclass[12pt]{article}

\usepackage{graphicx}
\usepackage{amsmath}
\usepackage{bm}
\usepackage{slashed}
\usepackage{amsfonts}

\newcommand\pubnumber{}
\newcommand\pubdate{\today}

\def\support{\footnote{zhang.5676@osu.edu. Speaker.}}

\def\Title#1{\begin{center} {\Large #1 } \end{center}}
\def\Author#1{\begin{center}{ \sc #1} \end{center}}
\def\Address#1{\begin{center}{ \it #1} \end{center}}

\newcommand\pubblock{\rightline{\begin{tabular}{l} \pubnumber\\
         \pubdate  \end{tabular}}}
\newenvironment{Abstract}{\begin{quotation}  }{\end{quotation}}
\newenvironment{Presented}{\begin{quotation} \begin{center}
             PRESENTED AT\end{center}\bigskip
      \begin{center}\begin{large}}{\end{large}\end{center} \end{quotation}}
\def\Acknowledgements{\bigskip  \bigskip \begin{center} \begin{large}
             \bf ACKNOWLEDGEMENTS \end{large}\end{center}}

\def\beq{\begin{equation}}
\def\eeq#1{\label{#1}\end{equation}}
\def\eeqn{\end{equation}}

\def\beqa{\begin{eqnarray}}
\def\eeqa#1{\label{#1}\end{eqnarray}}
\def\eeqan{\end{eqnarray}}

\let\bar=\overbar

\def\Dslash{\not{\hbox{\kern-4pt $D$}}}
\def\dslash{\not{\hbox{\kern-2pt $\del$}}}

\def\msb{{\bar{\ssstyle M \kern -1pt S}}}

\newcommand{\cc}{{Q\bar{Q}}}

\newcommand{\state}[4]{{^{#1}\hspace{-0.6mm}{#2}_{#3}^{[#4]}}}

\newcommand\CScSa{\state{3}{S}{1}{1}}

\newcommand\COaSz{\state{1}{S}{0}{8}}

\newcommand\COcSa{\state{3}{S}{1}{8}}

\newcommand\COcPj{\state{3}{P}{J}{8}}

\begin{document}
\begin{titlepage}
\pubblock

\Title{QCD factorization for high $p_T$ heavy quarkonium production}
\vfill
\Author{Yan-Qing Ma\footnote{yqma@pku.edu.cn}}
\Address{School of Physics and State Key Laboratory of Nuclear Physics and
Technology, Peking University, Beijing 100871, China\\
Collaborative Innovation Center of Quantum Matter,
Beijing 100871, China }
\vfill
\Author{Jian-Wei Qiu\footnote{jqiu@bnl.gov}}
\Address{Physics Department,
Brookhaven National Laboratory, Upton, NY 11973, USA\\
C.N. Yang Institute for Theoretical Physics and Department of Physics and Astronomy,
Stony Brook University, Stony Brook, NY 11794, USA }
\vfill
\Author{George Sterman\footnote{george.sterman@stonybrook.edu}}
\Address{C.N. Yang Institute for Theoretical Physics and Department of Physics and Astronomy,
Stony Brook University, Stony Brook NY 11794 USA }
\vfill
\Author{Hong Zhang\support}
\Address{Department of Physics,
The Ohio State University Columbus, OH 43210, USA }
\vfill
\begin{Abstract}
In this talk, we present the QCD factorization formula for heavy quarkonium production at large $p_T$ 
with factorized leading-power and next-to-leading power contributions in the $1/p_T$ expansion.
We show that the leading order analytical calculations in this QCD factorization approach can reproduce 
effectively the full next-to-leading order numerical results derived using non-relativistic QCD (NRQCD) 
factorization formalism.  We demonstrate that the next-to-leading power contributions are crucial to the description of the channels that are the most relevant for the rate as well as polarization of $J/\psi$ production 
at current collider energies.  
\end{Abstract}
\begin{Presented}
\vspace{-0.2cm}
The 7th International Workshop on Charm Physics (CHARM 2015)\\
Detroit, MI, 18-22 May, 2015
\end{Presented}
\end{titlepage}
\def\thefootnote{\fnsymbol{footnote}}
\setcounter{footnote}{0}
%

\section{Introduction}

Since the discovery of the $J/\psi$, heavy quarkonia, with their clearly separated multiple momentum scales, have been serving as ideal systems to test our understanding of QCD bound states and their hadronization processes. Unfortunately, a theoretically and phenomenologically satisfying framework is still not achieved for heavy quarkonium production, to explain their yields and the polarizations. The problem is more acute with the recently discovered $XYZ$-mesons, since the production of these exotic mesons requires good understanding of the production of conventional heavy quarkonia \cite{Brambilla:2010cs,Bodwin:2013nua}.

So far the most phenomenologically successful model for heavy quarkonium production is based on the non-relativistic QCD (NRQCD) factorization \cite{Caswell:1985ui,Bodwin:1994jh}, which factorizes the production cross section into the production of a heavy quark pair in different non-relativistic states, multiplied by the transition for the pair to transmute into the observed quarkonium. The production of the heavy quark pair is effectively perturbative and is organized in powers of $\alpha_s$ and $v$, the relative velocity of the heavy quark in the pair's rest frame, while the corresponding transition rate is nonperturbative and is represented by a set of NRQCD long-distance matrix elements (LDMEs). If the factorization is correct to all orders in $\alpha_s$ and powers in $v$, these LDMEs will be universal, i.e. process independent. Once the values of these LDMEs are extracted from one set of data, they can be used to explain and predict other data.

However, the current NRQCD factorization formalism is far from perfect in describing data on the heavy quarkonium production at high $p_T$.  The fixed-order NRQCD calculation suffers from large high-order corrections, due to the large enhancement in the powers of $p_T^2/m_Q^2$ and $\log(p_T^2/m_Q^2)$-type logarithms from high order contributions in powers of $\alpha_s$, where $m_Q$ is the mass of heavy quark \cite{Kang:2014tta}.
Although the most phenomenologically important contribution is assumed to be included in the NLO NRQCD calculation \cite{Ma:2010yw,Ma:2010jj}, the existence of large logarithm may potentially undermine the convergence of $\alpha_s$ expansion at large $p_T\gg m_Q$. The lack of convergence may also be the reason for the long-standing heavy quarkonium polarization puzzle \cite{Butenschoen:2011ks,Chao:2012iv,Gong:2012ug} and the recently-realized heavy-quark-spin-symmetry-violation puzzle \cite{Butenschoen:2014dra, Han:2014jya, Zhang:2014ybe}.

At large $p_T$, the unstable perturbative series in the NRQCD factorization approach is due to the existence of another small parameter $m_Q^2/p_T^2$, which is not included in the power counting of NRQCD effective theory. A satisfying framework for heavy quarkonium production at large $p_T$ must treat $m_Q^2/p_T^2$ in a systematic manner.

Recently, a new QCD factorization formalism has been proposed to study heavy quarkonium production at large $p_T$ \cite{Kang:2014tta, Kang:2011mg,Kang:2011zza, Fleming:2012wy, Fleming:2013qu, Kang:2014pya}. In this formalism, the production cross section is expanded in powers of $1/p_T$.  It was argued to all orders in $\alpha_s$ that the dominant leading-power (LP) terms, as well as the next-to-leading power (NLP) terms, can be systematically factorized into the perturbatively calculable hard parts for producing a single parton (or a heavy quark pair at NLP), convoluted with corresponding single parton (or heavy quark pair) fragmentation functions (FFs) to the observed heavy quarkonium.  Other than those suppressed by even higher powers in $1/p_T$, all nonperturbative contributions to the cross sections are included in these FFs, whose scale dependence is determined by a closed set of evolution equations with perturbatively calculable evolution kernels. By solving the evolution equations, large perturbative $\log(p_T^2/m_Q^2)$-type logarithms can be resummed to all orders in powers of $\alpha_s$. Because of the systematic treatment of the powers of $p_T^2/m_Q^2$ and $\log(p_T^2/m_Q^2)$, the QCD factorization approach may converge faster in the perturbative expansion in powers of $\alpha_s$ than the NRQCD factorization.

\section{QCD factorization for heavy quarkonium\\ production}

In the new QCD factorization approach, the production cross section of a heavy quarkonium $H$ with momentum $p$ at a large transverse momentum $p_T$ in the lab frame is expanded in a power series of $1/p_T$ \cite{Kang:2014tta,Kang:2011mg,Kang:2011zza,Kang:2014pya}
\begin{align}\label{eq:pQCDfac}
\begin{split}
E_p \frac{d\sigma_{A+B\to H+X}}{d^3p}(p)
&\approx
\sum_f\int\frac{dz}{z^2}D_{f\to H}(z;m_Q)
E_c\frac{d\hat{\sigma}_{A+B\to f(p_c)+X}}{d^3p_c}\Big(p_c=\frac{1}{z}\hat{p}\Big)
\\
&\hspace{0cm}
+\sum_{[Q\bar{Q}(\kappa)]}\int\frac{dz}{z^2}\, \frac{d\zeta_1\, d\zeta_2}{4}\, {\cal{D}}_{[Q\bar{Q}(\kappa)]\to H}(z,\zeta_1,\zeta_2;m_Q)
\\
&\hspace{1.5cm}
 \times
 E_c\frac{d\hat{\sigma}_{A+B\to[Q\bar{Q}(\kappa)](p_c)+X}}{d^3 p_c}
(P_Q,
P_{\bar{Q}};
P'_Q,
P'_{\bar{Q}}),
\end{split}
\end{align}
where the first (second) term on the right-hand side is the LP (NLP) contribution, and the error of this factorization formula is suppressed by the power of $\mathcal{O}(m_Q^4/p_T^4)$ and higher. The LP (NLP) contribution is  factorized into a sum of convolutions of the production of a single parton $f$ (a heavy quark pair in state $\kappa$) in the hard collision and the single-parton (heavy-quark-pair) fragmentation functions to all orders of $\alpha_s$ \cite{Kang:2014tta}. For the LP term, the parton $f$ can be gluon, light quark, heavy quark or their anti-particles, while the $\cc$ state $\kappa$ can be vector, axial vector or tensor, either color-singlet or color-octet for the NLP term. In Eq.~(\ref{eq:pQCDfac}), $z$ is the longitudinal momentum fraction of the fragmenting parton(s), which is taken by the heavy quarkonium;  $\zeta_1$ and $\zeta_2$ are the relative longitudinal momentum of the $\cc$-pair in the amplitude and that in the complex conjugate of the amplitude, respectively, and can be different.

In Eq.~(\ref{eq:pQCDfac}), the short-distance partonic hard parts $d\hat{\sigma}$ could be systematically calculated in powers of $\alpha_s$ (we need to convolute with parton distribution functions (PDFs) if $A$ and/or $B$ is a hadron). The FFs $D_{f\to H}(z;m_Q,\mu_F)$ and ${\cal{D}}_{[Q\bar{Q}(\kappa)]\to H}(z,\zeta_1,\zeta_2;m_Q,\mu_F)$ are intrinsically nonperturbative, but process independent, universal functions. However, their dependence on factorization scale $\mu_F$ is determined by a closed set of evolution equations \cite{Kang:2014tta},
\begin{align}
\begin{split}\label{eq:evo_1}
&\frac{\partial}{\partial\ln\mu_F^2} D_{f\to H}(z;m_Q,\mu_F)
=
\sum_{f'}
\int_{z}^1 \frac{dz'}{z'}
{D}_{f'\to H}(z';m_Q, \mu_F)\ \gamma_{f\to f'} (z/z',\alpha_s)
\\
&\hspace{3cm}
+
\frac{1}{\mu_F^2}\
\sum_{[Q\bar{Q}(\kappa')]}
\int_{z}^1 \frac{dz'}{z'} \int_{-1}^1 \frac{d\zeta'_1}{2}  \int_{-1}^1 \frac{d\zeta'_2}{2} \,
{\mathcal D}_{[Q\bar{Q}(\kappa')]\to H}(z', \zeta'_1,  \zeta'_2; m_Q, \mu_F)
\\
&\hspace{5cm}
\times
\gamma_{f\to [Q\bar{Q}(\kappa')]}({z}/{z'}, \zeta'_1, \zeta'_2) ,
\end{split}\\
\begin{split}\label{eq:evo_2}
&
\frac{\partial}{\partial{\ln\mu_F^2}}{\mathcal D}_{[Q\bar{Q}(\kappa)]\to H}(z, \zeta_1,\zeta_2;m_Q, \mu_F)
\\
&\hspace{1.5cm}
=
\sum_{[Q\bar{Q}(\kappa')]}
\int_{z}^1 \frac{dz'}{z'}
\int_{-1}^1 \frac{d\zeta'_1}{2}  \int_{-1}^1 \frac{d\zeta'_2}{2}\,
{\mathcal D}_{[Q\bar{Q}(\kappa')]\to H}(z', \zeta'_1, \zeta'_2;m_Q, \mu_F)
\\
&\hspace{2cm}
\times
\Gamma_{[Q\bar{Q}(\kappa)]\to [Q\bar{Q}(\kappa')]}(\frac{z}{z'}, \zeta_1, \zeta_2; \zeta'_1, \zeta'_2) ,
\end{split}
\end{align}
where the evolution kernels $\gamma's$ and $\Gamma's$ are process-independent and perturbatively calculable. The well-known DGLAP evolution kernels $\gamma_{f\to f'}$ are available to next-to-next-to-leading order in $\alpha_s$. The power-mixing evolution kernels $\gamma_{f\to [Q\bar{Q}(\kappa')]}$ were calculated in Ref.~\cite{Kang:2014tta}, and the heavy quark pair evolution kernels $\Gamma_{[Q\bar{Q}(\kappa)]\to [Q\bar{Q}(\kappa')]}$ have been recently calculated by two groups independently~\cite{Fleming:2013qu,Kang:2014tta}. If both $\kappa$ and $\kappa'$ are color singlet, the kernel $\Gamma_{[Q\bar{Q}(\kappa)]\to [Q\bar{Q}(\kappa')]}$ reduces to the well-known Efremov-Radyushkin-Brodsky-Lepage evolution kernel for exclusive processes~\cite{Lepage:1979zb,Efremov:1978rn}. By solving Eqs.~\eqref{eq:evo_1} and \eqref{eq:evo_2}, the perturbative logarithms $\log(p_T^2/m_Q^2)$ are summed to all orders.

\section{Calculation of fragmentation functions with\\ NRQCD factorization}

A set of input FFs at an initial scale $\mu_0$ is required for solving the evolution equations. If the input scale is chosen as $\mu_0\sim 2m_Q\gg \Lambda_{\rm QCD}$, it is natural to use NRQCD factorization to further factorize these input FFs~\cite{Ma:2013yla}

\begin{align}\label{eq:FFNR1}
\begin{split}
{ D}_{f \to H}(z; m_Q,\mu_0)
&=\hspace{-0.2cm}
\sum_{[\cc(n)]} \hat{d}_{[\cc(\kappa)]\to[\cc(n)]}(z; m_Q,\mu_0,\mu_\Lambda)
\langle \mathcal{O}_{[\cc(n)]}^{H}(\mu_\Lambda)\rangle,
\end{split}\\
\begin{split}\label{eq:FFNR2}
{\cal D}_{[\cc(\kappa)]\to H}(z,\zeta_1,\zeta_2; m_Q,\mu_0)
&=\hspace{-0.2cm}
\sum_{[\cc(n)]} \hat{d}_{[\cc(\kappa)]\to[\cc(n)]}(z,\zeta_1,\zeta_2;m_Q,\mu_0,\mu_\Lambda)
\langle \mathcal{O}_{[\cc(n)]}^{H}(\mu_\Lambda)\rangle,
\end{split}
\end{align}
where $\mu_0$ and $\mu_\Lambda$ are pQCD and NRQCD factorization scales, respectively.
The summation of intermediate $[\cc(n)]$ runs over all possible NRQCD states, which are labelled by spectroscopic notation $\state{{2S+1}}{L}{J}{1,8}$. The NRQCD LDMEs are ordered by powers of $v$, the relative velocity of the heavy quarkonium in its rest frame. In practice, the summation in Eqs.~\eqref{eq:FFNR1} and \eqref{eq:FFNR2} can be effectively truncated and only finite terms are kept on the right-hand side.

With the NRQCD factorization, all the $z$, $\zeta_1$ and $\zeta_2$ dependences of input FFs can be calculated perturbatively and given by the short-distance coefficients $\hat{d}$. The nonperturbative dynamics are absorbed in a few NRQCD LDMEs, which need to be determined by fitting experimental data. We calculated these short-distance coefficients to the first nontrivial order for all $S-$wave and $P-$wave polarization-summed heavy quarkonium states. Conventional dimensional regularization was applied to regularize both UV and IR divergences. All derived short-distance coefficients $\hat{d}$'s are IR-finite, and can be found in Refs.~\cite{Ma:2013yla, Ma:2014eja}.

Since the polarization is an important observable for exploring the production mechanism of heavy quarkonium \cite{Butenschoen:2011ks,Chao:2012iv,Gong:2012ug}, it is critically important to study the production of polarized heavy quarkonia in the framework of the QCD factorization formalism, which requires a set of polarized input FFs. To make the calculation compatible with conventional dimensional regularization, in Ref.~\cite{Ma:2015yka}, we define the $D-$dimensional polarized NRQCD LDMEs  by requiring that they preserve the same symmetry under rotation of angle $\pi$ about the pre-selected $z$-direction as their 4-dimensional counterparts. 
With the definition of $D$-dimensional NRQCD LDMEs, we calculate the short-distance coefficients in Eqs.~\eqref{eq:FFNR1} and \eqref{eq:FFNR2} to the first nontrivial order for all $S-$wave and $P-$wave polarized heavy quarkonium states. The obtained $\hat{d}$'s are all IR-finite, and can be found in Ref.~\cite{Ma:2015yka}.

\section{Comparison of LO QCD factorization with NLO\\ NRQCD factorization for $J/\psi$ production}

\begin{figure}[h!]
\centering
\includegraphics[height=2.7in]{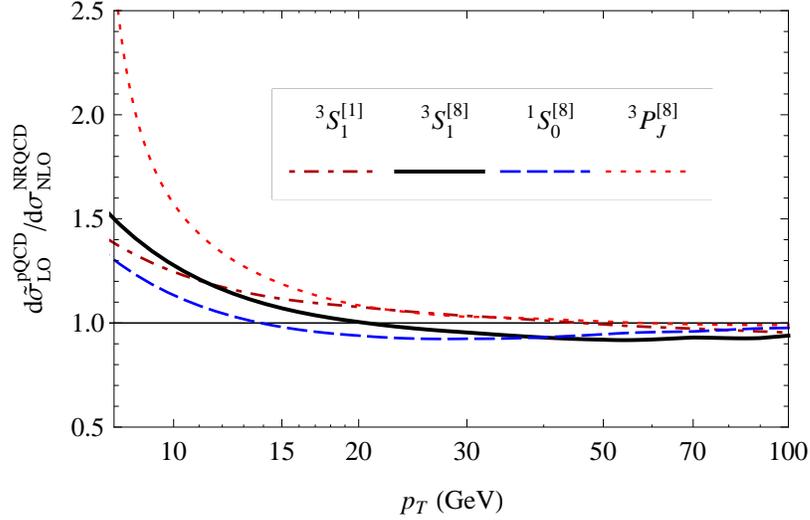}
\caption{Ratio of $J/\psi$ production rate from LO QCD factorization over that of NLO NRQCD calculation for four leading NRQCD channels. See the text for details.}
\label{fig:ratioNLONRQCD}
\end{figure}

As shown in Fig.~\ref{fig:ratioNLONRQCD}, our LO factorized QCD contributions to the production rate~\cite{Ma:2014svb} can almost reproduce the full NLO NRQCD calculation channel-by-channel for $p_T>10-15$ GeV.  The comparison in Fig.~\ref{fig:ratioNLONRQCD} demonstrates that the very complicated and numerically-evaluated results of NLO NRQCD calculations can be reproduced by the simple and fully analytic LO calculation of the QCD factorization approach for $p_T>10$ GeV, and clearly indicates that perturbative organization of the factorized power expansion is well suited to heavy quarkonium production at high $p_T$.  It also shows the importance of the NLP contribution.  Without it, as shown in Ref.~\cite{Bodwin:2014gia}, for example, the LP QCD factorization contribution can only reproduce NLO NRQCD results for the $\COcSa$ and $\COcPj$ channels at large $p_T$, not for the $\CScSa$ and $\COaSz$ channels.

\begin{figure}[h!]
\centering
\includegraphics[height=2.7in]{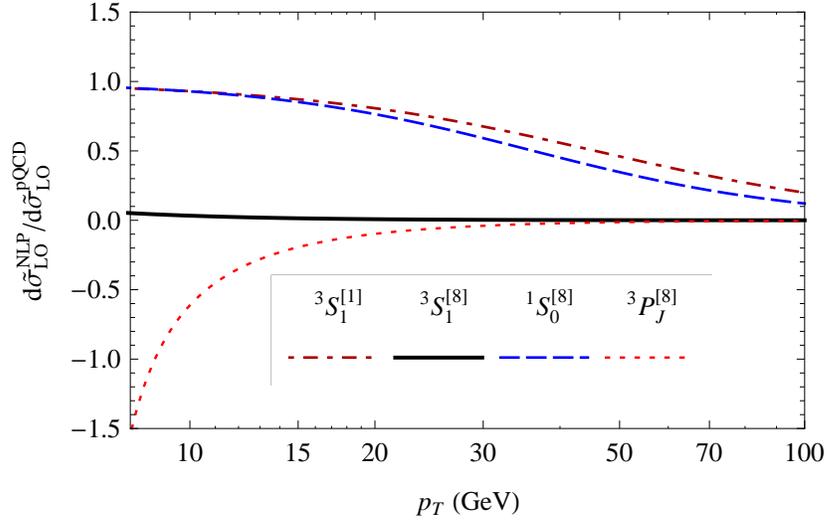}
\caption{Ratio of NLP contributions to total contribution in LO QCD for each channel. $d\tilde{\sigma}$ means we have a special choice for PDFs. See text for details.}
\label{fig:ratioNLP}
\end{figure}

To further illustrate the importance of NLP contributions, we plot the ratio of the NLP contribution to the total LO QCD contribution in Fig.~\ref{fig:ratioNLP} for each channel.  These ratios are independent of the values of the LDMEs.  Figure~\ref{fig:ratioNLP} clearly shows that NLP contributions are negligible for the $\COcSa$ channel over the full $ p_T$ range, and are small for the $\COcPj$ channel when $p_T>20$ GeV, beyond which it is below $10$ percent. On the other hand, the NLP contributions are crucial for $\COaSz$ and $\CScSa$ channels, which remain very significant in the production rate, even if $p_T$ approaches 100 GeV.  Since the FFs for a single active parton to fragment into a $\CScSa$ heavy quark pair, calculated in NRQCD, vanish for both LO and NLO, the two-loop gluon FF derived in Refs.~\cite{Braaten:1993rw,Bodwin:2012xc} was used for the LP contribution to the $\CScSa$ channel in Fig.~\ref{fig:ratioNLP}.

In the above comparison with NLO NRQCD calculations, we did not include the evolution of FFs.  A complete LO QCD calculation should include the evolution of FFs using the LO evolution kernels given in Ref.\ \cite{Kang:2014tta} and input FFs calculated in NRQCD factorization at NLO \cite{Ma:2013yla,Ma:2014eja}, and a set of updated NRQCD LDMEs by fitting the data.  From its consistency with the existing NLO NRQCD results, and the control through evolution of its higher order corrections, we expect the LO factorized QCD power expansion to do a better job interpreting existing data on heavy quarkonium production at collider energies, while a consistent NLO contribution is within our reach.  Also, because the LP  $\COcSa$ and $\COcPj$ channels, which produce predominantly transversely polarized heavy quarkonia, appear not to be dominant \cite{Chao:2012iv,Bodwin:2014gia,Faccioli:2014cqa}, heavy quarkonium production at current collider energies is strongly influenced by the $\COaSz$ channel, and is more likely to be unpolarized.



\Acknowledgements
This work was supported in part by the U. S. Department of Energy under contracts No. DE- AC02-05CH11231 and No. DE-AC02-98CH10886, and the National Science Foundation under Grants No. PHY- 0969739 and No. PHY-1316617.


\end{document}